\documentstyle[jkas]{article}

\beginpage{125}
\endpage{134}
\year{2011}\volume{44}\month{August}
\runningauthor {H. A. N. LE ET AL.}
\runningtitle{SPECTRAL LIBRARY OF LATE-TYPE STELLAR TEMPLATES}
\month{August} \year{2011} \volume{44} \issueno{4}
\beginpage{125}\endpage{134}

\begin{document}

\title{MEDIUM RESOLUTION SPECTRAL LIBRARY OF LATE-TYPE STELLAR TEMPLATES IN NEAR-INFRARED BAND}

\author{Huynh Anh Nguyen Le$^{1}$, Wonseok Kang$^{1}$, Soojong Pak$^{1}$,  Myungshin Im$^{2}$, Jeong-Eun Lee$^{1}$, Luis C. Ho$^{3}$, Tae-Soo Pyo$^{4}$, and Daniel T. Jaffe$^{1,5}$}

\address{$^1$ School of Space Research, Kyung Hee University, 1 Seocheon-dong, Giheung-gu, Yongin, Gyeonggi-do 446-701, Korea
\\ {\it E-mail : huynhanh7@khu.ac.kr, wskang@khu.ac.kr, soojong@khu.ac.kr, and jeongeun.lee@khu.ac.kr}}
\address{$^2$ Department of Physics and Astronomy, Seoul National University, Seoul, 151-747, Korea
\\ {\it E-mail : mim@astro.snu.ac.kr}}
\address{$^3$ The Observatories of the Carnegie Institution for Science, 813 Santa Barbara Street, Pasadena, CA 91101, USA
\\ {\it E-mail : lho@obs.carnegiescience.edu }}
\address{$^4$ Subaru Telescope, National Astronomical Observatory of Japan, 650 North Aohoku Place, Hilo, HI 96720, USA
\\ {\it E-mail : pyo@subaru.naoj.org }}
\address{$^5$ Department of Astronomy, University of Texas at Austin, Austin, TX 78712, USA
\\ {\it E-mail : dtj@astro.as.utexas.edu }}
\address{\normalsize{\it (Received January 07, 2011; Revised June 30, 2011; Accepted July 25, 2011)}}

\offprints{W. Kang}

\abstract{We present medium resolution (R $=$ $5000-6000$) spectra in the near-infrared band, $1.4-1.8$ $\mu$m, for template stars in G, K, and M types observed by the echelle spectrometer, IRCS, at the SUBARU 8.2 m telescope. The identification of lines is based on the spectra of Arcturus (K2 III) in the literature. We measured the equivalent of widths and compared our results to those of Meyer {\rm et al.} (1998). We conclude that our spectral resolution (R $=$ $6000$) data can investigate more accurately the properties of lines in stellar spectra. The library of the template stellar spectra in ASCII format are available for download on the World Wide Web.}

\maketitle


\section{INTRODUCTION}

The near-infrared (near-IR) band of the stellar spectra have been poorly studied, even though they are useful windows to examine natures of various astronomical sources, e.g., the physical and chemical processes taking place in stellar atmosphere (Heras {\rm et al.} 2002). In addition, such stellar templates are potentially useful for studying the relation between active galactic nuclei (AGNs) and their host galaxy properties, providing reference lines for measuring the velocity dispersions (e.g., Riffel {\rm et al.} 2008).

Current efforts in compiling libraries of template stars in G, K, and M types have been performed at low to moderate spectral resolution (R $\leq$ 3000; Wallace \& Hinkle 1997; Meyer {\rm et al.} 1998; Rayner {\rm et al.} 2009; Pickles {\rm et al.} 1998). Recently, new instruments on the 8-m class telescopes provide medium and high resolution spectra of R $=$ $5000-20000$. It is now important to provide stellar spectral library that matches the resolution of these newer instruments.

In this paper, we present the near-IR H-band spectra ($1.4 - 1.8$ $\mu$m) of G, K and M type stars. The spectra are taken with moderate resolution (R of $=$ $5000-10000$), improving the spectral resolution of the previous works, or broadening the wavelength coverage. Section 2 presents the samples, the observations, and the data reduction processes. In Section 3, the results and discussions of the spectra are given. We calculate and analyze equivalent widths (EWs) of the line and suggest a way to estimate the effective temperature for giant stars from EWs. Summary and future works are presented in Section 4.

\section{OBSERVATIONS AND DATA REDUCTIONS}

\subsection{Sample}

We select the bright template stars ($H < 5$ mag) in spectral classes (G, K, and M) in the luminosity class of III. The giant stars are expected to be the dominant stellar populations of bulges of AGN host galaxies. Table~\ref{sample} shows the spectral types and near-IR magnitudes of our samples.

\begin{table}[t]
\caption{Sample selection}
\centering
\scriptsize
\doublerulesep2.0pt
\renewcommand\arraystretch{1.5}
\begin{tabular}{llccc}
\hline
\hline
Object  &  Spectral   & J$^{\rm a}$  &  H$^{\rm a}$  &   K$^{\rm a}$    \\
 Name   &    Type      &     (mag) & (mag) & (mag) \\
\hline
HD39357 & A0V & 4.789$\pm$0.286 & 4.734$\pm$0.264 & 4.487$\pm$0.016 \\
HD105388 & A0V & 7.343$\pm$0.021 & 7.390$\pm$0.024 & 7.359$\pm$0.027 \\
HD64938 & G8III & 4.892$\pm$0.244 & 4.192$\pm$0.234 & 4.049$\pm$0.276 \\
HD148287 & G8III & 4.635$\pm$0.212 & 4.013$\pm$0.194 & 3.957$\pm$0.202 \\
HD55184 & K0III & 4.204$\pm$0.328 & 3.473$\pm$0.276 & 3.437$\pm$0.298 \\
HD155500 & K0III & 4.377$\pm$0.272 & 3.760$\pm$0.228 & 3.706$\pm$0.194 \\
HD52071 & K2III & 5.041$\pm$0.232 & 4.295$\pm$0.198 &  \nodata \\
HD146084 & K2III & 4.614$\pm$0.250 & 3.823$\pm$0.200 & 3.741$\pm$0.230 \\
HD122675 & K2III & 4.316$\pm$0.302 & 3.538$\pm$0.222  & 3.244$\pm$0.344 \\
HD154610 & K5III & 3.943$\pm$0.268 & 3.165$\pm$0.200 & 2.896$\pm$0.252 \\
HD76010 & M0III & 4.266$\pm$0.268 & 3.332$\pm$0.198 & 3.095$\pm$0.278 \\
NSV3729 & M0III & 2.055$\pm$0.278 & 1.216$\pm$0.190 & 1.144$\pm$0.200 \\
\hline
\end{tabular}
\label{sample}
\begin{tabnote}
\vskip10pt
\hskip-30pt $^{\rm a}$ Magnitudes in SIMBAD (http:$\slash$$\slash$simbad.u-strasbg.fr$\slash$).\\
\hskip-90pt Reference: 2003yCat2246... (Cutri) \\
\end{tabnote}
\end{table}
%
\begin{table}[!tb]
\caption{Observation log}
\centering
\doublerulesep2.0pt
\scriptsize
\renewcommand\arraystretch{1.5}
\resizebox{0.5\textwidth}{!}{
\begin{tabular}{l@{  }l@{}c@{}c@{}c@{}c@{}c@{}c}
\hline
\hline
Date  & Object &  Echelle &  Slit    & R &  Total      &  Seeing  & Airmass  \\
(UT)  &  Name  &  Setting &  Width   &   &  Exposure  &          &          \\
      &        &          & (arcsec) &   & (sec)       & (arcsec) &          \\
\hline
2003 Feb 11   &       HD39357  &    $H+$    &   0.3 & 10000  &  $4 \times 20$  &    0.80   & 1.06    \\
     ...    &             HD122675   &   $H+$  &  0.3 &  10000  &    $4 \times 12$  &  0.80 & 1.33  \\
     ...    &          NSV3729      &  $H+$  &   0.3 &  10000  &    $4 \times 10$  & 0.80 & 1.06    \\[0.5ex]
2004 Apr 3 &          HD105388  &     $H-$   &  0.6 &  5000 &    $4 \times 20$ &  0.07 & 1.08  \\
            &            ...      &   $H+$     &  0.6  & 5000  &   $4 \times 20$  & 0.07 & 1.08   \\
    ...   &          HD55184    &    $H+$   &  0.6 &  5000 &       $4 \times 6$ & 0.30 &  1.04  \\
    ...   &             ...      &    $H-$    &   0.6 & 5000 &     $4 \times 6$ & 0.30  & 1.04  \\
    ... &            HD64938  &      $H-$   & 0.6 &  5000 &        $4 \times 10$  &  0.30 & 1.04   \\
     ... &               ...  &      $H+$    &  0.6 &  5000 &      $4 \times 10$ & 0.30 & 1.04  \\
    ... &               HD148287 &      $H-$   & 0.6 &  5000 &     $4 \times 10$  & 0.14 & 1.08 \\
    ... &                ...     &       $H+$   &  0.6 & 5000 &    $4 \times 10$ & 0.14 & 1.08  \\
    ... &               HD155500  &     $H-$   &  0.6  &   5000 &  $4 \times 10$  & 0.14 & 1.06 \\
    ... &                ...      &      $H+$   &    0.6 &  5000 & $4 \times 10$  & 0.14 & 1.06 \\[0.5ex]
2004 Apr 4 &       HD52071  &       $H-$  &  0.6&  5000 &       $4 \times 6$  & 0.40 & 1.04   \\
    ... &                ...    &       $H+$  &    0.6 & 5000 &   $4 \times 6$ &   0.40 & 1.04 \\
    ... &              HD76010  &       $H-$   &  0.6 &  5000 &    $4 \times 5$ & 0.40  & 1.01 \\
    ... &                ...    &        $H+$  &   0.6 &  5000 &   $4 \times 5$ &  0.40 &  1.01  \\
    ... &              HD154610 &       $H-$   &  0.6 & 5000 &     $4 \times 5$ & 0.36 & 1.30  \\
    ... &                ...    &        $H+$ &  0.6 &   5000 &      $4 \times 5$ &  0.36  & 1.30  \\
    ... &             HD146084 & $H+$ &   0.6 &  5000 &    $4 \times 10$ &  0.36 & 1.30    \\
    ... &               ...    &       $H-$ & 0.6 &  5000 &       $4 \times 10$  &  0.36 &  1.30  \\[0.5ex]
\hline
\end{tabular}
\label{observation}
}
\end{table}

\subsection{Observation}

The observations were performed at the SUBARU 8.2 m telescope using the IRCS (Tokunaga {\rm et al }. 1998b; Kobayashi {\rm et al}. 2000) on 2003 Feb 11 UT and 2004 April 3 and 4 UT. Table \ref{observation} shows logs of the observations.

In the 2003 observations, we observed two template stars, K2 III (HD122675) and M0 III (NSV3729). We also took an A0 V standard star (HD39357) to correct the telluric absorption lines in the target spectra. The weather conditions were clear and the natural seeing was 0.8 arcsec. The slit width was 0.3 arcsec whose spectral resolving power is 10000. The position angle of the slit was 0 deg. The echelle settings of the observations were in $H+$ band covering the wavelength range of $1.47-1.82$ $\mu$m. The simple frame exposure time was about 10 sec for each target.

The standard A0 V star and the template stars were observed in the nod-on-slit ({\it ABBA}) mode in which the target is taken in two positions, $A$ and $B$, along the slit.

In 2004, we observed eight template stars, G8 III (HD64938 and HD148287), K0 III (HD55184 and HD155500), K2 III (HD52071 and HD146084), K5 III (HD154610), and M0 III (HD76010). The standard star A0 V (HD105388) was also observed. The slit width was 0.6 arcsec whose spectral resolving power is 5000. The echelle was set to cover the whole H-bands in the standard $H-$ and $H+$ settings. Other instrument settings and observing modes were the same as in 2003.

\subsection{Data Reductions}

\begin{figure*}[!thp]
\centering
\includegraphics[width=75ex]{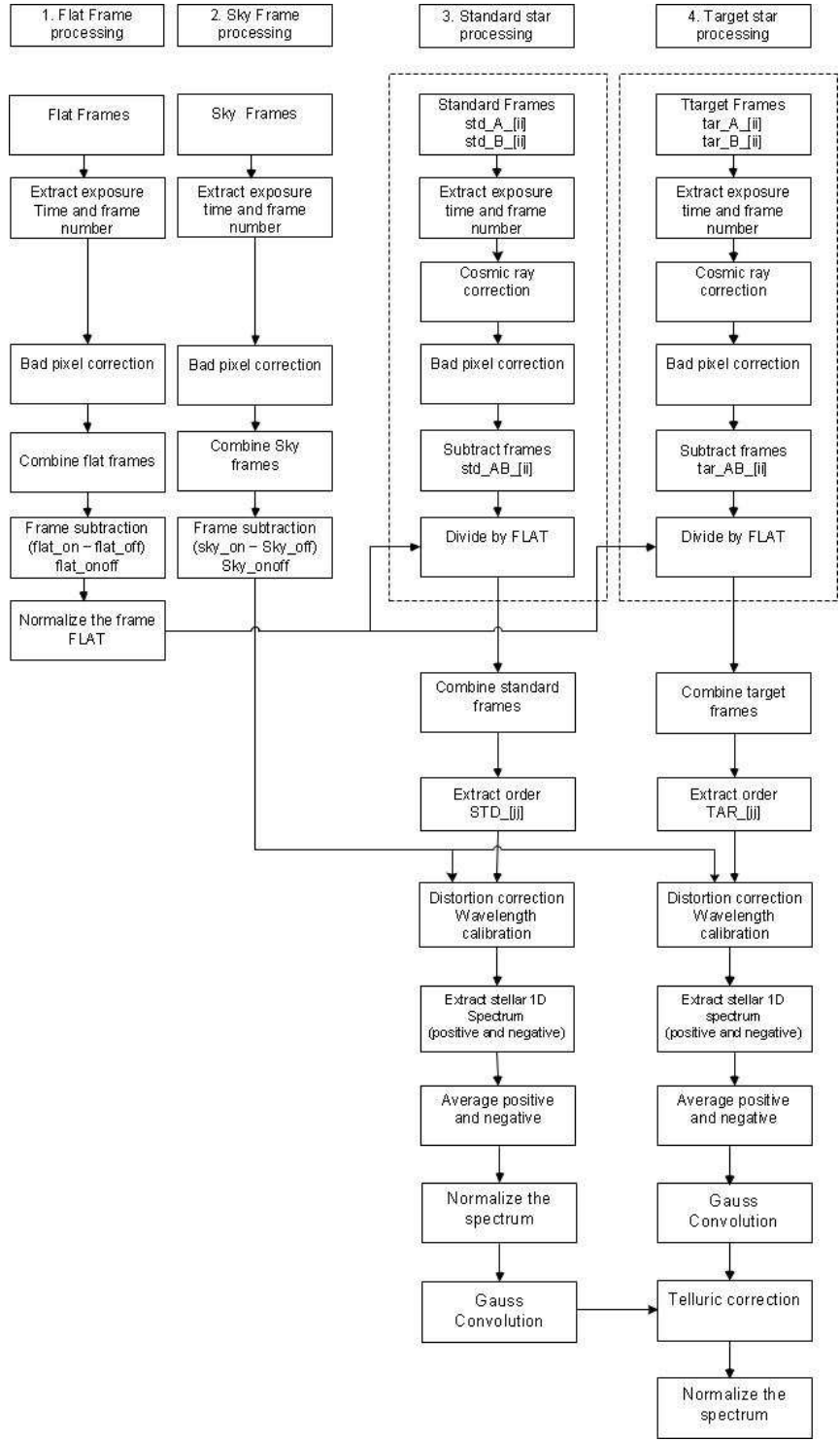}
\caption{Data Reduction Processes.}
\label{data}
\end{figure*}

Data reduction processes were done by {\it IRAF} \footnote{{\it IRAF} (Image Reduction and Analysis Facility) is distributed by the National Optical Astronomy Observatories (NOAO) which is operated by the Association of Universities for Research in Astronomy (AURA), Inc. under cooperative agreement with the National Science Foundation.} followed by the method described in Pyo (2002). In addition, Python language was used in some tasks to derive the final spectra of targets. Fig. \ref{data} shows the data reduction sequences.

We first made a combined dark frame. All the dark-subtracted data images were corrected for cosmic rays and bad pixels. We used a bad pixel mask (BadPixel070502.fits) provided by the SUBARU IRCS team. Cosmic rays were identified using the detection threshold of 1000 in {\it COSMICRAYS} within a $5 \times 5$ box area. Flat-field correction was done by using normalized flat field images. We also derived the order extraction solutions for echellogram images. This task will be explained in the reduction processes of the standard star. We used telluric OH emission lines as references for wavelength calibrations and geometric distortions (Rousselot {\rm et al}. 2000).

\begin{figure}[!t]
\centering
\includegraphics[width=45ex]{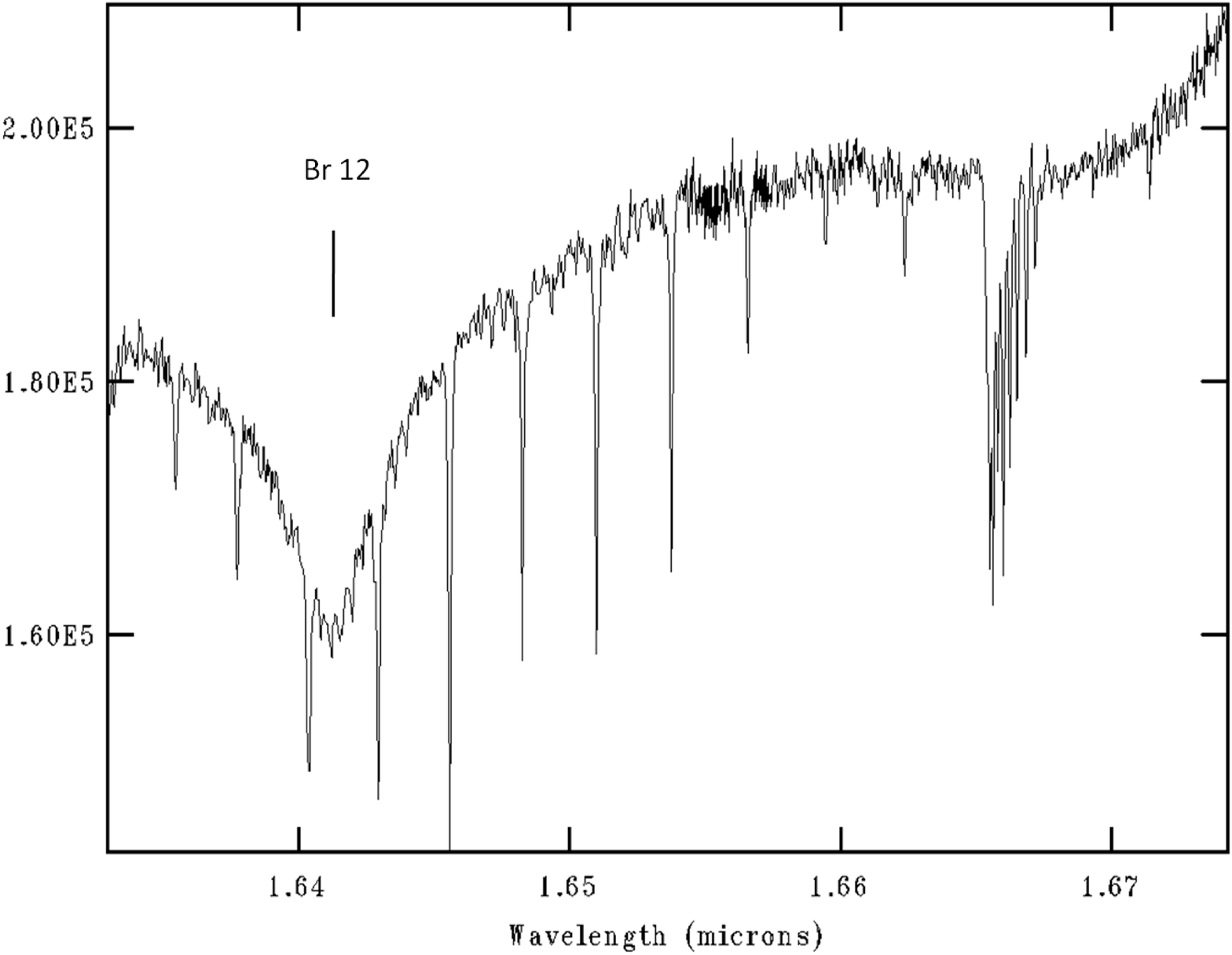}
\includegraphics[width=45ex]{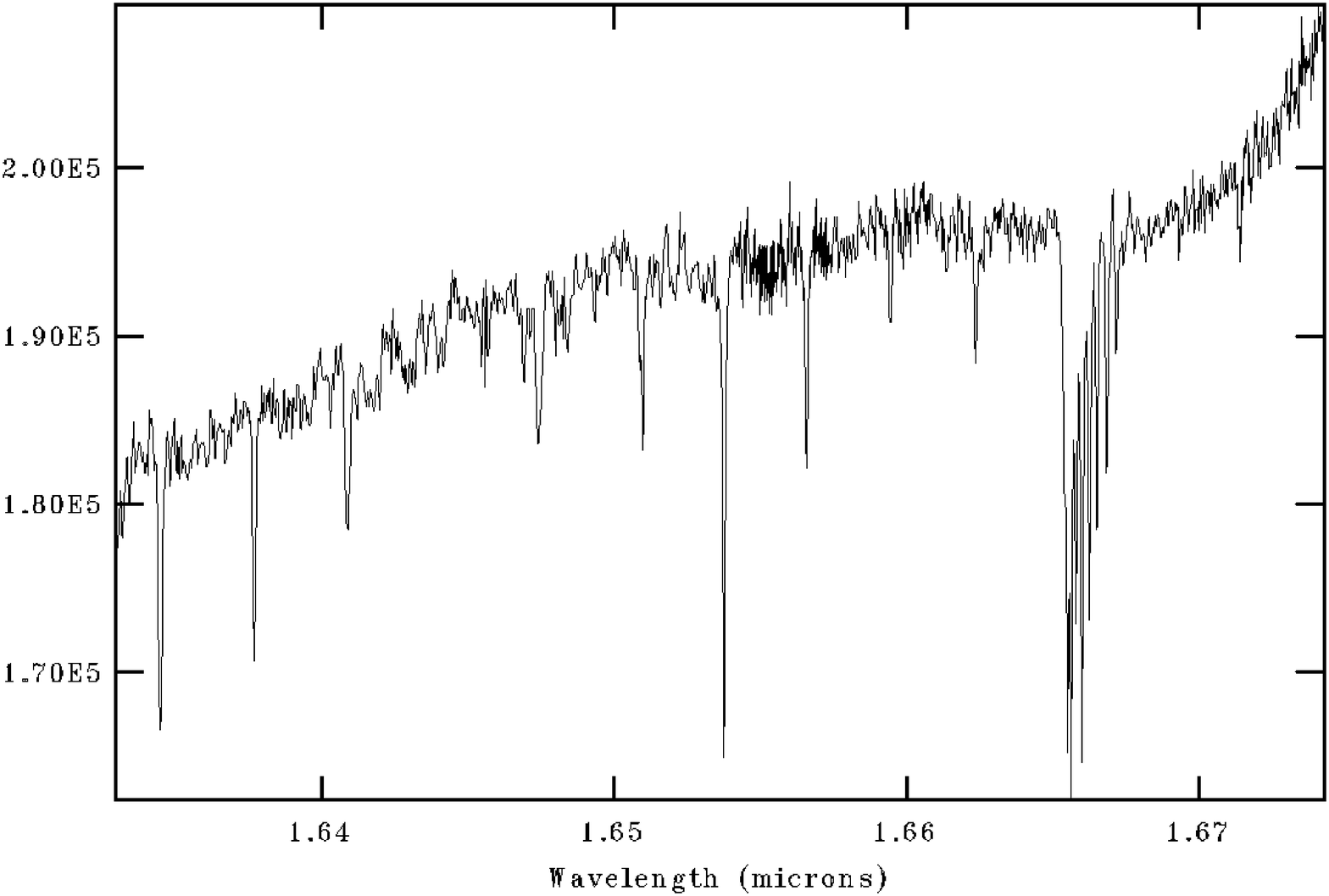}
\caption{Sample spectra of HD39357 (A0 V). The top panel shows a Bracket line Br12 at 1.6412 $\mu$m, and the bottom panel after Br12 correction processes.}
\label{bracketA0V}
\end{figure}

\begin{figure}[!t]
\centering
\includegraphics[width=45ex]{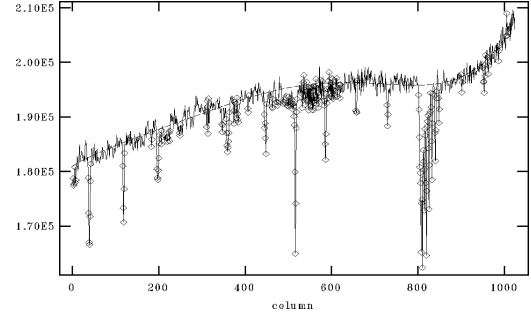}
\caption{Sample spectrum of HD39357 (A0 V) during continuum task in IRAF.}
\label{continuumA0V}
\end{figure}

The telluric absorption lines in the stellar spectra is corrected by dividing for the standard star A0 V spectra. The spectra of standard star needs to be corrected for the Bracket lines from the stellar atmosphere. To do this step, we fitted a Gaussian profile to a Bracket absorption line using the {\it SPLOT} task in {\it IRAF}. Fig. \ref{bracketA0V} shows the spectrum of A0 V star with Bracket lines before and after correction. The {\it CONTINUUM} task is used for normalization of the flux (see Fig. \ref{continuumA0V}).

\begin{figure}[!t]
\centering
\includegraphics[width=45ex]{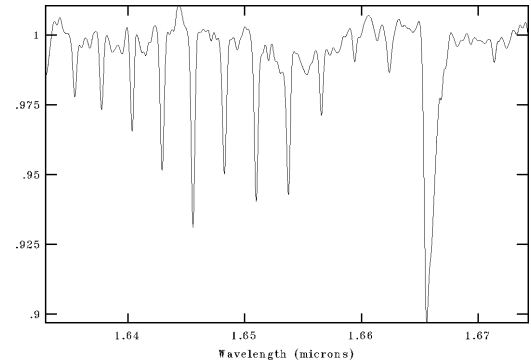}
\caption{Sample spectrum of HD39357 (A0 V) after GAUSS task in IRAF.}
\label{gaussA0V}
\end{figure}

In the case of the slit width of 0.6 arcsec, the spectral resolution of the point source depends on the seeing size, while the unresolved spectral lines of the extended sources like sky background or galaxies show a box-shaped profile. In order to match the spectral resolutions of the telluric OH emissions, the standard stars, and the template stars, we convolved the spectra by using the {\it GAUSS} process with sigma = 2.942 to make FWHM = 8 pixels (1 pixel = $3.94 \times 10^{-5}$ $\mu$m). After the spectral convolution process, the resulting resolution is between $5000$ and $6000$. Fig. \ref{gaussA0V} shows the spectra after the {\it GAUSS} process.

\begin{figure}[!t]
\centering
\includegraphics[width=55ex]{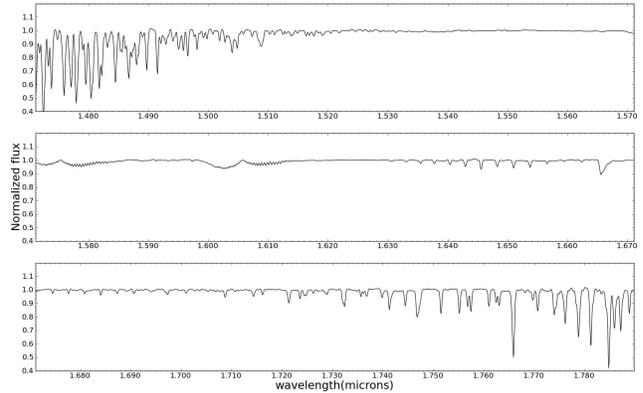}
\caption{Spectra of A0 V standard stars, from combining HD39357 and HD10538.}
\label{A0V}
\end{figure}

Finally, after getting the final spectrum of each aperture from {\it IRAF}, we made a laced average spectrum for A0 V standard star using Python language. In this program, we concatenated all the aperture spectra to one spectrum in H-band, and made average spectrum of $H+$ and $H-$ settings. Fig. \ref{A0V} shows the final spectrum of A0 V standard star whose wavelength has a step of 0.0001{$\mu$m}. Note that the signal-to-noise ratio is very high (S/N $\approx$ 100).

The data reduction processes of template stars are the same as those of the standard stars, except the extraction of 1D spectrum for the distortion corrections, and Bracket line corrections. By dividing for the A0 V star spectra, we corrected the telluric absorption lines in the template spectra. Finally, we used the {\it CONTINUUM} task to normalize the flux in the spectra.

The Doppler shift in the stellar spectra due to Earth$'$s motions was corrected by using heliocentric velocity which is obtained from the {\it RVCORRECT} task. We also corrected the Doppler shift of the proper motions by checking a prominent stellar absorption line.


\section{RESULTS AND DISCUSSIONS}

\subsection{Template Spectra}

\begin{figure*}[!ht]
\centering
\includegraphics[width=120ex]{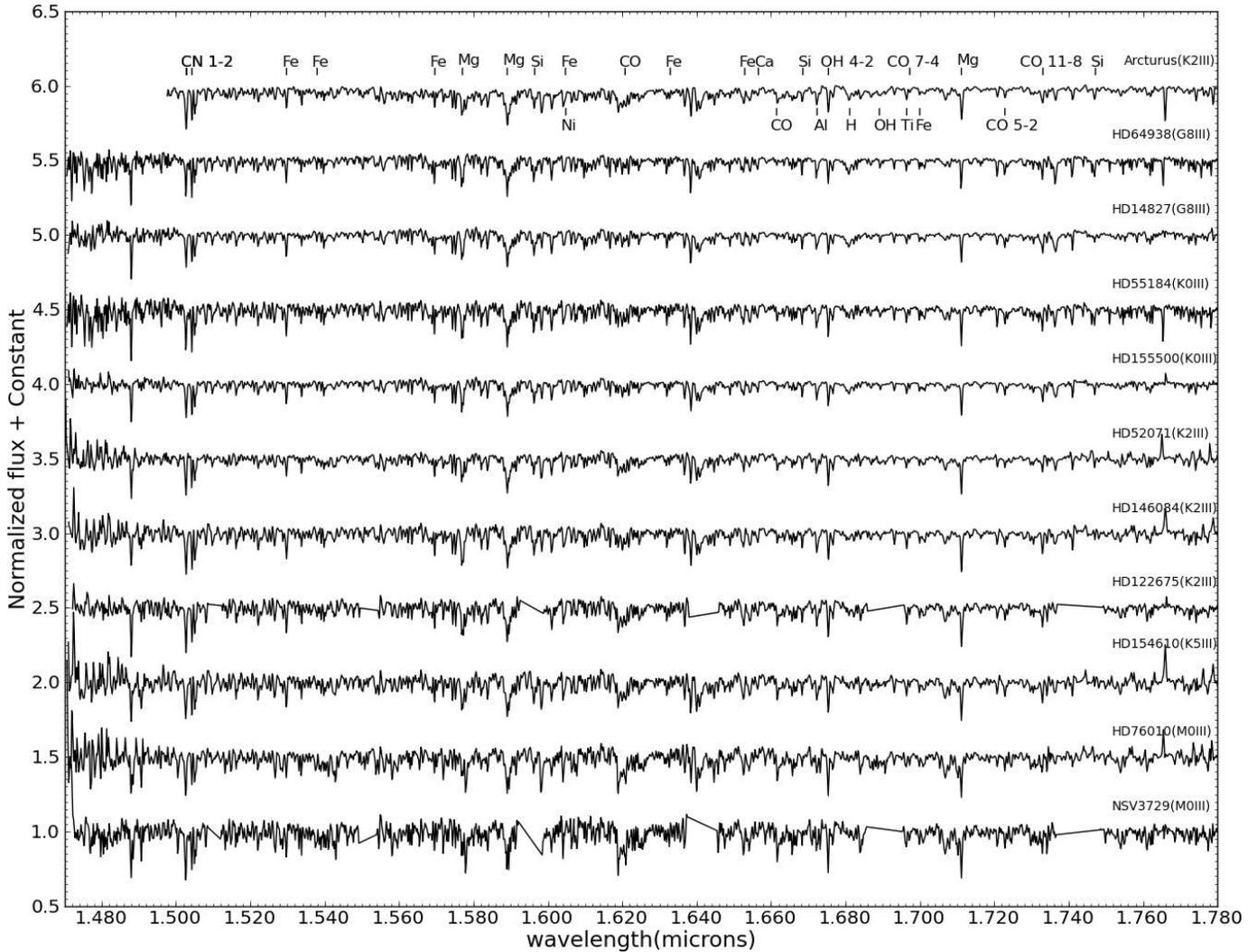}
\caption{Spectra of the template stars in G-K-M types. All of the identified features in the spectra are indicated at the Arcturus (K2 III) spectrum.}
\label{compare}
\end{figure*}

Fig. \ref{compare} shows reduced template spectra in this work and spectrum of Arcturus (K2 III) from Hinkle \& Wallace \rm et al.(1995) which is convoluted by using the {\it GAUSS} task to convert the spectral resolution from R $=$ $100000$ to R $=$ $6000$. We can see that most of the identified features are very similar to each other. Table \ref{identify_Hinkle} lists the identified lines in our data.

Note that some normalized flux values of the spectra are above unity. Even though they are not noise of the spectra, they are just regarded as noise during the {\it CONTINUUM} task. In addition, the emission lines at $\lambda$ $<$ 1.48 $\mu$m and $\lambda$ $>$ 1.76 $\mu$m are not real lines since these regions are on the boundary of H-band where the telluric absorption lines are too deep to be subtracted.

\begin{figure}[t]
\centering
\includegraphics[width=35ex]{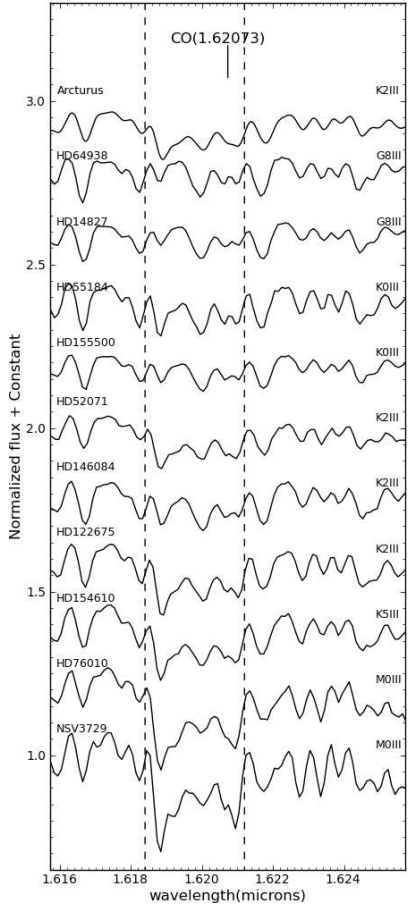}
\caption{CO (1.62073 $\mu$m) line for Arcturus and G-K-M type stars. The dash lines show the ranges of EW measurements. CO (1.62073 $\mu$m) is potential for measuring the velocity dispersion in the central region of QSO host galaxies.}
\label{CO_162073}
\end{figure}

Our obtained H-band spectra cover the the broad absorption $\nu$ $=$ $3-6$ $^{12}$CO band head at 1.62073 $\mu$m. This broad absorption feature is very useful for kinematic extraction. This absorption feature has a robust comparison between stellar spectra and host galaxy spectra (McConnell \rm et al. 2011). We plan to use this CO band head for measuring the velocity dispersion in the central region of QSO host galaxies. Fig. \ref{CO_162073} shows the broad absorption $\nu$ $=$ $3-6$ $^{12}$CO band head at 1.62073 $\mu$m in our spectra of G8-M0 stars, including Arcturus (K2 III).

\begin{table}[th]
\caption{Identified lines in the reduced template star spectra}
\begin{center}
\centering
\scriptsize
\doublerulesep2.0pt
\renewcommand\arraystretch{1.5}
\begin{tabular}{cccc}
\hline
\hline
Line$^{\rm a}$	& Wavelengths  & Features  & EW    \\
                 & ($\mu$m)     &           & measurement \\
\hline
CN 1-2 	 & 1.50284  & Strong &  No   \\
CN 1-2 	& 1.50430   & Strong &  No   \\
Fe	& 1.52973       & Strong &  No   \\
Fe	& 1.53795       & Weak &  No   \\
Fe	& 1.56961       & Strong &  No   \\
Mg & 1.57701        & Strong &  Yes    \\
Mg	& 1.58905       & Strong &  No    \\
Si & 1.59644        & Strong &  Yes    \\
Fe & 1.60471        & Strong &  No    \\
Ni &  1.60481       & Weak &  No    \\
CO & 1.62073       & Broad & Yes   \\
Fe 	& 1.63289       & Weak &  No   \\
Fe	& 1.65290       & Strong &  No    \\
Ca	& 1.65656       & Weak &  No    \\
CO & 1.66147        & Broad & Yes   \\
Si & 1.66853        & Strong &  Yes   \\
Al & 1.67235        & Strong &  Yes   \\
OH 4-2   &	1.67538 & Broad & No   \\
H & 1.68111         & Broad & Yes  \\
OH & 1.68909        & Narrow & Yes \\
Ti	& 1.69644       & Strong  & No   \\
CO 7-4 	 & 1.69727  & Narrow & No  \\
Fe	& 1.69994       & Weak & No  \\
Mg & 1.71113        & Strong & Yes   \\
CO 5-2 	& 1.72284   & Weak & No  \\
CO 11-8 	& 1.73307 & Strong & No \\
Si & 1.74717 & Weak & No  \\[0.5ex]
\hline
\end{tabular}
\end{center}
\label{identify_Hinkle}
\begin{tabnote}
\hskip10pt $^{\rm a}$ Identified lines from Hinkle \& Wallace (1995).\\
\end{tabnote}
\end{table}

\subsection{Equivalent Widths}

From the identified lines of our spectra based on the referent lines of Arcturus (K2 III) spectra (Hinkle \& Wallace \rm et al. 1995), we measured the equivalent widths (EWs) of 9 features which are prominent in the spectra on A-M spectral types (Meyer \rm et al. 1998, hereafter M98). The formula for estimating the EWs and the variance, $\sigma_{EW}$, are given by

\begin{equation}
EW = \int_{\lambda _{min}}^{\lambda _{max}}\left \{ 1-F(\lambda ) \right \}d\lambda
\end{equation}

\begin{equation}
\sigma_{EW}= \Delta \lambda \sqrt{ < {\sigma _F}^2 >  }
\end{equation}

Where $F(\lambda)$ is the normalized flux, and $\sigma _F$ is the root mean square value (RMS). Table $4$ shows the results of EWs in units of wave number ($cm^{-1}$). The effective temperature in the table are taken from Tokunaga (1998a).

\begin{figure*}[th!]
\centering
\includegraphics[width=74ex]{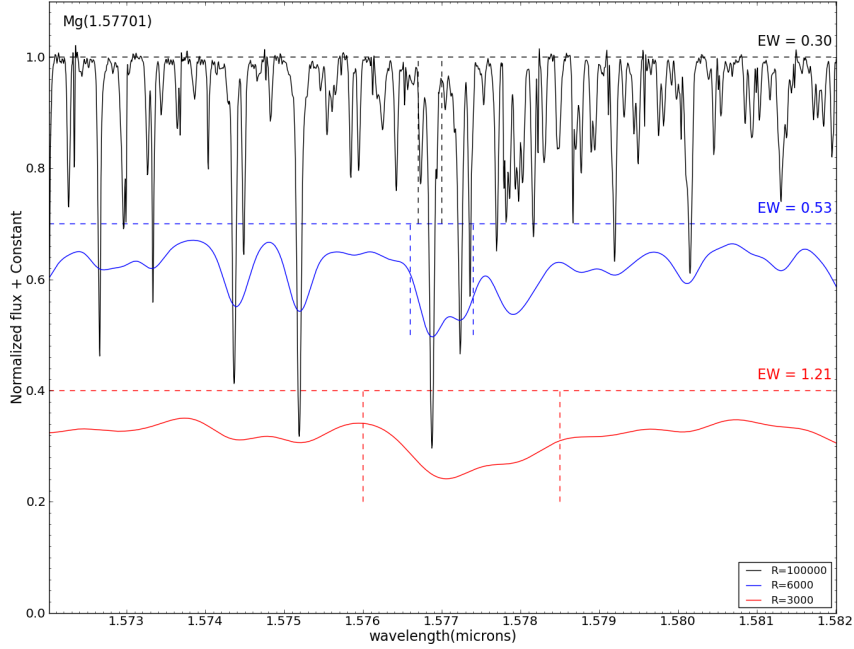}
\caption{Mg (1.57701 $\mu$m) spectra in Arcturus (Hinkle \& Wallace 1995) with resolutions of R $=$ 100000 (top), R $=$ 6000 (middle), and R $=$ 3000 (bottom). The horizontal dark, blue and red dash lines show the continua. The vertical dark (1.5767 $\mu$m-1.5770 $\mu$m), blue (1.5766 $\mu$m-1.5774 $\mu$m), and red (1.5760 $\mu$m-1.5785 $\mu$m) dash lines show the width ranges ($\lambda _{min}$, $\lambda _{max}$) which use for calculating the EWs. More lines can be seen in the spectrum of R $=$ 6000 compare with that of R $=$ 3000. Because the contaminated lines can not be separated of R $=$ 6000 and R $=$ 3000, the EW values are larger than those of R $=$ 100000.}
\label{Mg15Arcturusline}
\end{figure*}

\begin{figure*}[th!]
\centering
\includegraphics[width=74ex]{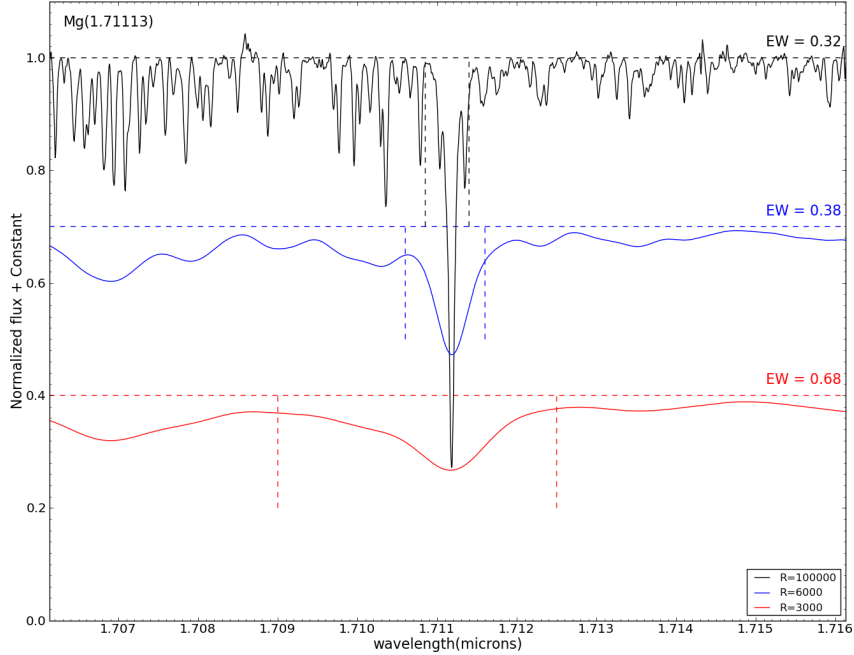}
\caption{Mg (1.71113 $\mu$m) line is a single line, and very sensitive to the temperature of giant stars. The wavelength range of EW measurement for R $=$ 100000 and R $=$ 6000 is similar to each other, so the difference of EWs between R $=$ 100000 and R $=$ 6000 is negligible. On the other hand, in the case of R $=$ 3000, it is difficult to define the wavelength range and to measure the EW of this line.}
\label{Mg17Arcturusline}
\end{figure*}

In order to investigate the effect of spectral resolution, we convolved the high resolution spectra (R $=$ $100000$) of Arcturus (Hinkle \& Walllace 1995) to make the resolutions correspond to our spectra of R $=$ $6000$ and to the M98 spectra of R $=$ $3000$. Fig. \ref{Mg15Arcturusline} and \ref{Mg17Arcturusline} show two Mg lines (1.57701 $\mu$m and 1.71113 $\mu$m) of Arcturus with different spectral resolution. Table \ref{ArcturusEW} shows the results of EW measurements from the Arcturus spectrum of each resolution in wave number units ($cm^{-1}$). Assuming that $\sigma_F$ is 0.1 in the Arcturus spectrum, we calculated the errors of EWs from the formula (2). The Mg (1.57701 $\mu$m) line is severely contaminated with other lines, so the EW results of R $=$ 6000 and R $=$ 3000 are different from that of R = 100000. And the EW of Mg (1.71113 $\mu$m) line for the spectrum of R $=$ 6000 is very similar to that of R $=$ 100000, while that of R $=$ 3000 shows large difference from the EW result of R $=$ 100000. From a comparison with EWs in spectra with a different resolution, we can assert that our EW results in the spectrum with R $=$ 6000 are more accurate than those of M98 with R $=$ 3000.

\begin{figure*}[th!]
\includegraphics[width=27ex]{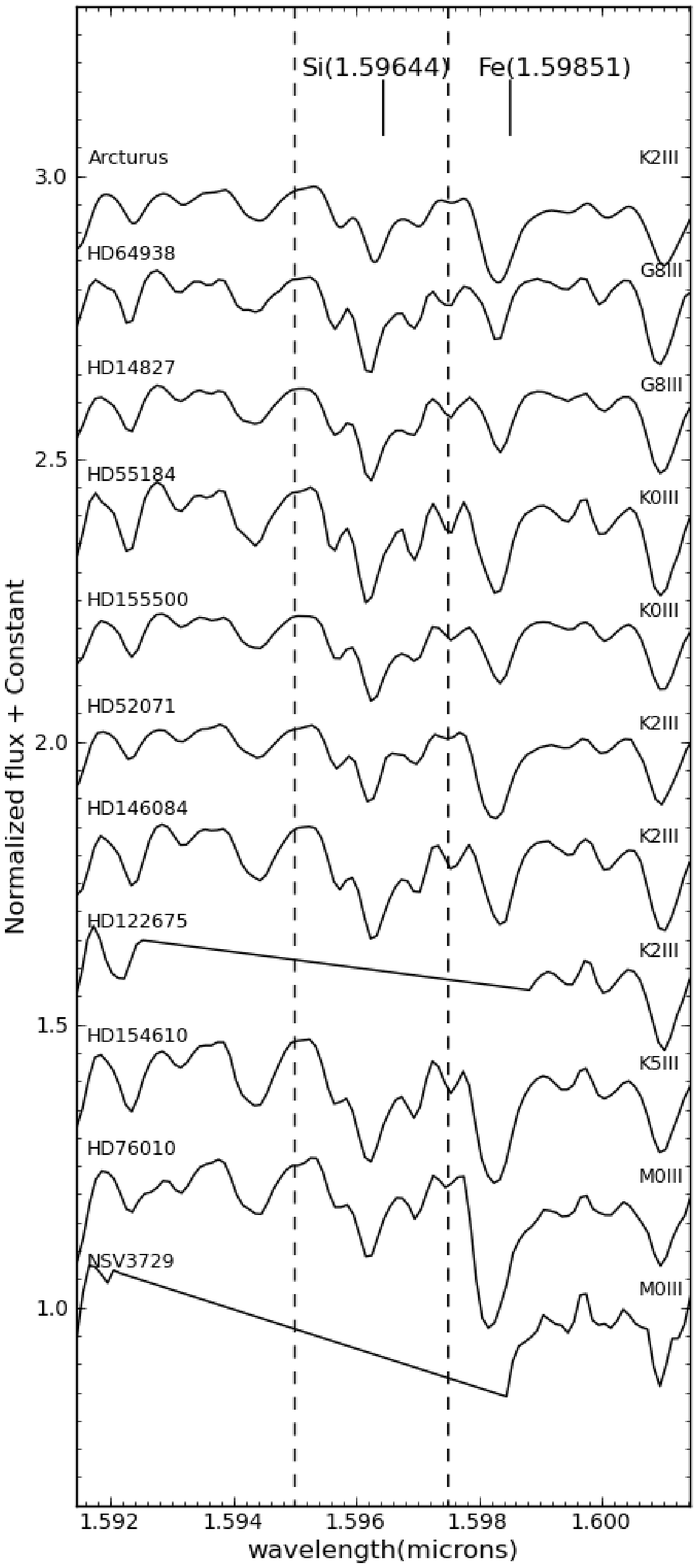}
\includegraphics[width=27ex]{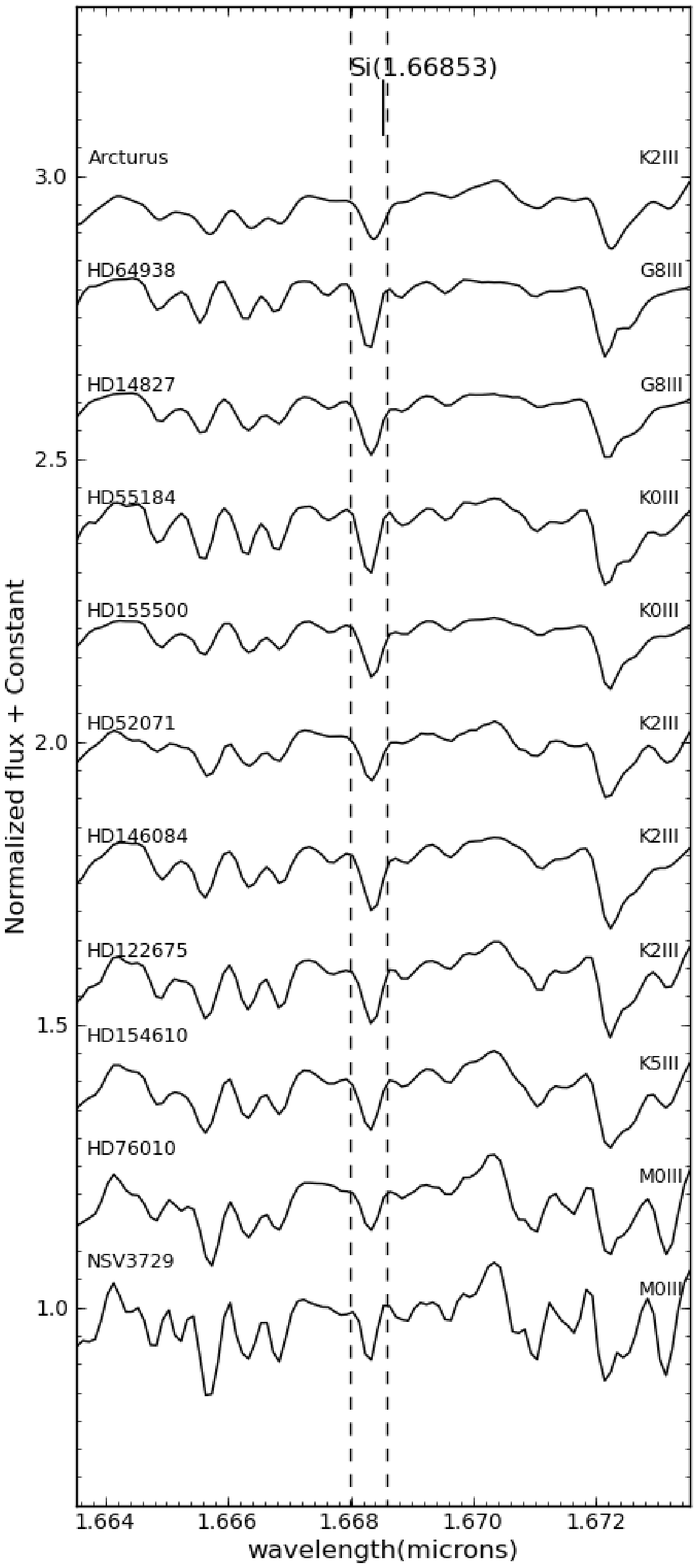}
\includegraphics[width=28ex]{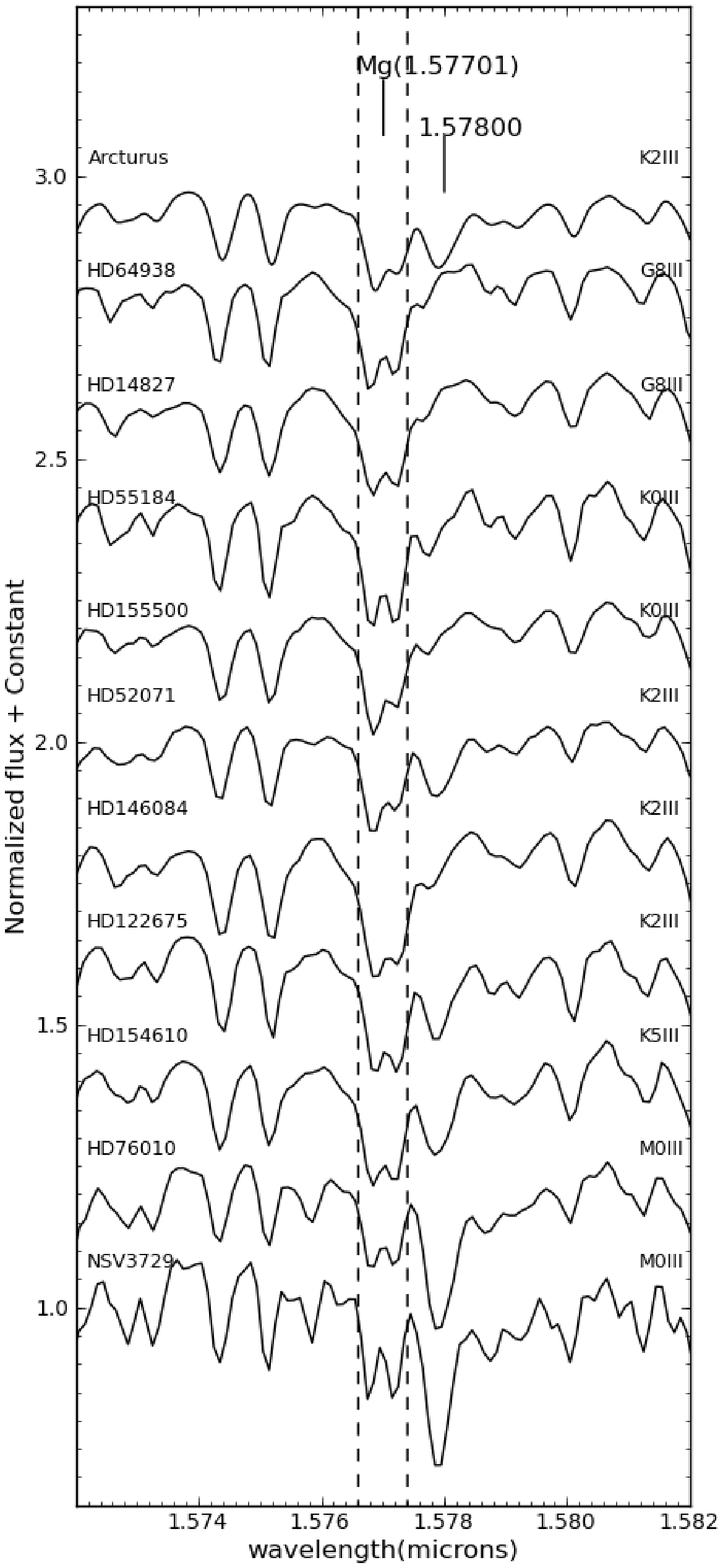}
\includegraphics[width=27.7ex]{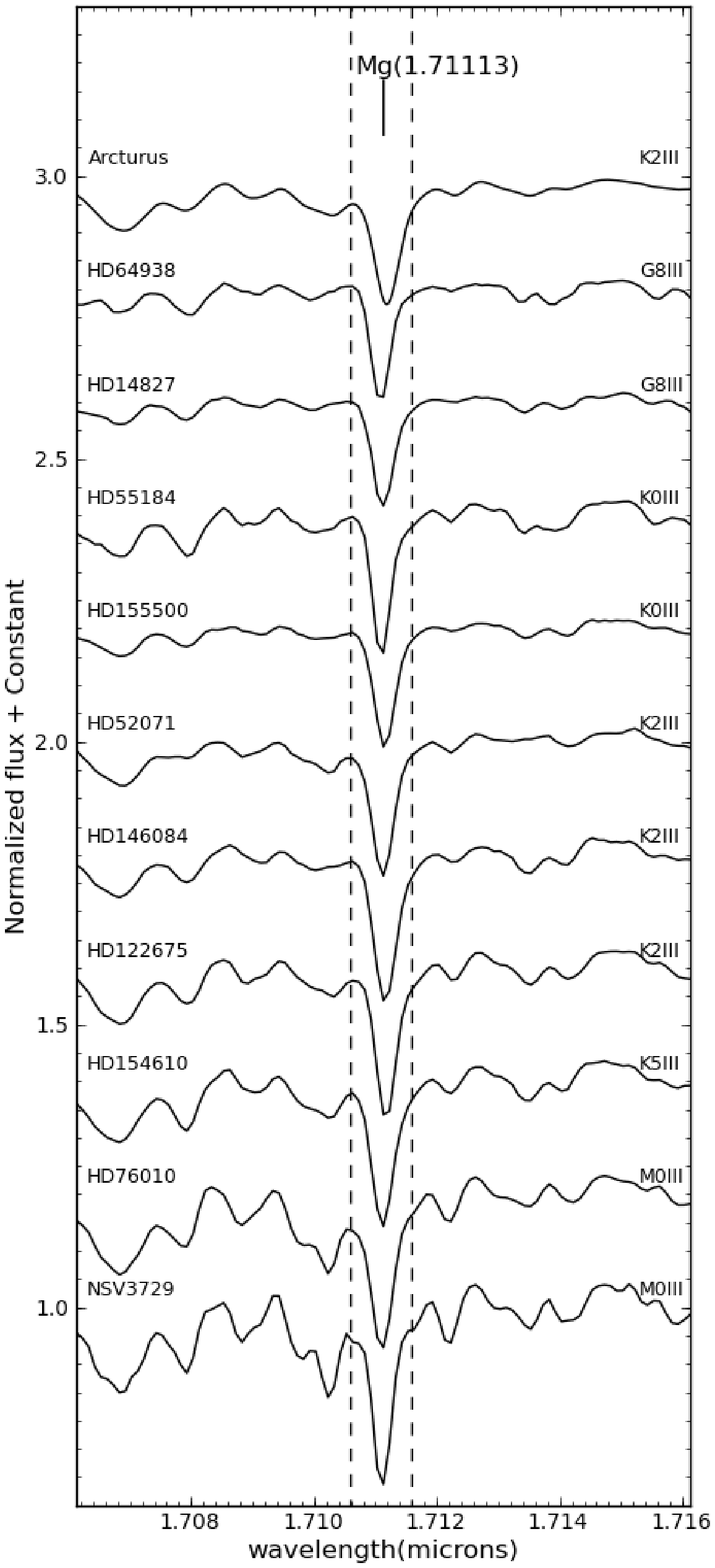}
\caption{The Si (1.59644 $\mu$m), Si (1.66853 $\mu$m), Mg (1.57701 $\mu$m), Mg (1.71113 $\mu$m) lines for Arcturus and G-K-M type stars. The top spectrum is from the convoluted one of Arcturus with the same resolution of the other spectra. The dark dash lines show the width range for calculating the EWs. Note that the Si (1.59644 $\mu$m) line is contaminated with the strong Fe (1.59851 $\mu$m) line, and the Mg (1.57701 $\mu$m) line might be also contaminated with the nearby line (1.57800 $\mu$m).}
\label{allline}
\end{figure*}


%
%

\subsection{Equivalent Width vs. Temperature}

\begin{table*}[tb!]
\begin{center}
\caption{Equivalent widths$^{\rm a}$ \label{equivalents}}
\scriptsize
\doublerulesep2.0pt
\renewcommand\arraystretch{1.5}
\resizebox{0.99\textwidth}{!}{%
\begin{tabular}{llcccccccccc}
\hline
\hline
Object Name &  Spectral   & Teff$^{\rm b}$  &  Mg[5843]        & OH[5920]          &  H[5948]         & Al[5980]        & Si[5993]        & CO[6019]        & CO[6170]        & Si[6264]        &  Mg[6341]    \\
            &      Type           & (K)      & (1.71113 $\mu$m)  & (1.68909 $\mu$m)   & (1.68111 $\mu$m)  & (1.67235 $\mu$m) & (1.66853 $\mu$m) & (1.66147 $\mu$m) & (1.62073 $\mu$m) & (1.59644 $\mu$m) & (1.57701 $\mu$m)  \\
\hline
HD64938     & G8III           & 4960     & 0.26$\pm$0.02    & 0.07$\pm$0.01     & 0.58$\pm$0.02   & 0.23$\pm$0.02    & 0.12$\pm$0.02   & 0.06$\pm$0.01   & 0.31$\pm$0.04   & 0.46$\pm$0.02   & 0.43$\pm$0.03   \\
HD148287    & G8III           & 4960     & 0.27$\pm$0.03    & 0.08$\pm$0.01     & 0.56$\pm$0.02   & 0.21$\pm$0.02    & 0.12$\pm$0.02   & 0.05$\pm$0.02   & 0.29$\pm$0.03   & 0.41$\pm$0.02   & 0.43$\pm$0.02   \\
HD55184     & K0III           & 4810     & 0.33$\pm$0.04    & 0.09$\pm$0.01     & 0.25$\pm$0.03   & 0.25$\pm$0.03    & 0.11$\pm$0.03   & 0.22$\pm$0.03   & 0.63$\pm$0.06   & 0.37$\pm$0.03   & 0.50$\pm$0.03   \\
HD155500    & K0III           & 4810     & 0.32$\pm$0.03    & 0.05$\pm$0.01     & 0.39$\pm$0.01   & 0.20$\pm$0.02    & 0.10$\pm$0.02   & 0.09$\pm$0.01   & 0.39$\pm$0.02   & 0.37$\pm$0.02   & 0.43$\pm$0.02   \\
HD52071     & K2III           & 4500     & 0.39$\pm$0.03    & 0.12$\pm$0.01     & 0.15$\pm$0.01   & 0.18$\pm$0.02    & 0.08$\pm$0.02   & 0.22$\pm$0.02   & 0.75$\pm$0.04   & 0.24$\pm$0.02   & 0.37$\pm$0.02   \\
HD146084    & K2III           & 4500     & 0.42$\pm$0.03    & 0.09$\pm$0.01     & 0.29$\pm$0.01   & 0.24$\pm$0.02    & 0.12$\pm$0.02   & 0.16$\pm$0.02   & 0.60$\pm$0.04   & 0.36$\pm$0.02   & 0.57$\pm$0.02   \\
HD122675    & K2III           & 4500     & 0.41$\pm$0.03    &   \nodata         & 0.21$\pm$0.01   & 0.19$\pm$0.02    & 0.12$\pm$0.01   & 0.41$\pm$0.02   & 0.97$\pm$0.03   &   \nodata       & 0.47$\pm$0.02   \\
HD154610    & K5III           & 3980     & 0.42$\pm$0.03    & 0.11$\pm$0.01     & 0.09$\pm$0.02   & 0.24$\pm$0.03    & 0.10$\pm$0.02   & 0.30$\pm$0.03   & 0.98$\pm$0.05   & 0.25$\pm$0.02   & 0.48$\pm$0.02   \\
HD76010     & M0III           & 3820     & 0.48$\pm$0.03    & 0.29$\pm$0.01     & 0.04$\pm$0.02   & 0.18$\pm$0.03    & 0.06$\pm$0.01   & 0.51$\pm$0.03   & 1.39$\pm$0.06   & 0.11$\pm$0.03   & 0.32$\pm$0.03   \\
NSV3729     & M0III           & 3820     & 0.49$\pm$0.04    &   \nodata         & 0.09$\pm$0.02   & 0.21$\pm$0.03    & 0.09$\pm$0.02   & 0.69$\pm$0.04   & 1.55$\pm$0.05   &   \nodata       & 0.35$\pm$0.04   \\
\hline
\end{tabular}}
\end{center}
\begin{tabnote}
 \hskip0pt $^{\rm a}$ Values in units of $cm^{-1}$.\\
 \hskip0pt $^{\rm b}$ Effective temperature taken from Tokunaga (1998a).\\
\end{tabnote}

\end{table*}

\begin{table}[thp]
\begin{center}
\caption{Equivalent widths$^{\rm a}$ of Arcturus$^{\rm b}$ (K2 III)}
\scriptsize
\doublerulesep2.0pt
\renewcommand\arraystretch{1.5}
\begin{tabular}{lcc}
\hline
\hline
Spectral Resolution & Mg[6341]   &  Mg[5843]   \\
  &   Mg(1.57701 $\mu$m)         & Mg(1.71113 $\mu$m)           \\
\hline
$R=100000$      &  0.30$\pm$0.02               & 0.32$\pm$0.03                 \\
$R=6000$        &  0.53$\pm$0.04               & 0.38$\pm$0.04                 \\
$R=3000$        &  1.21$\pm$0.07               & 0.68$\pm$0.07                 \\
\hline
\end{tabular}
\end{center}
\begin{tabnote}
\hskip0pt $^{\rm a}$ Values in units of $cm^{-1}$.\\
\hskip-20pt $^{\rm b}$ Spectra from Hinkle \& Wallace (1995).\\
\end{tabnote}
\label{ArcturusEW}
\end{table}

\begin{figure*}[ht!]
\centering
\includegraphics[width=90ex]{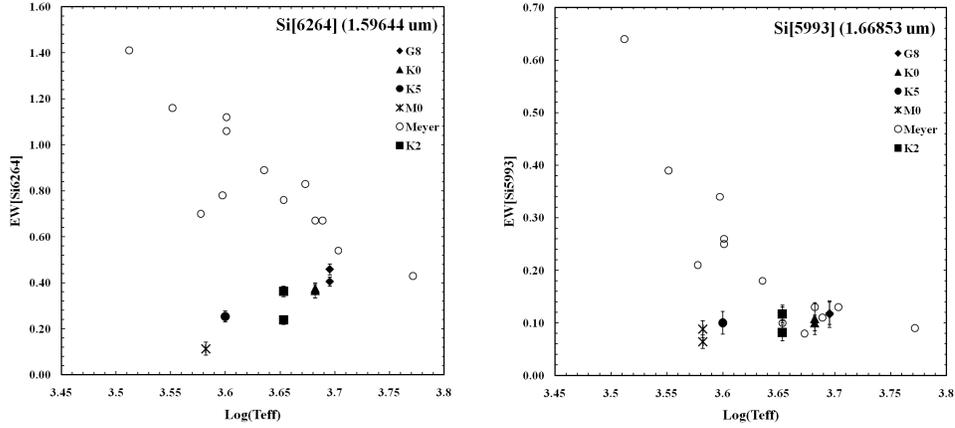}
\caption{EWs of Si (1.59644 $\mu$m) and Si (1.66853 $\mu$m). Our results are smaller than M98 results because of the different spectral resolution. The trends of Si (1.59644 $\mu$m) line are very different compared to M98. The Si (1.59644 $\mu$m) line is contaminated with other lines, and in M98 results, they may have used the width range which is too wide and consequently contaminated with other lines including the strong Fe (1.59851 $\mu$m) line. This strong Fe line makes our results very different from that of M98.}
\label{Si2lines}
\end{figure*}

\begin{figure*}[ht!]
\centering
\includegraphics[width=90ex]{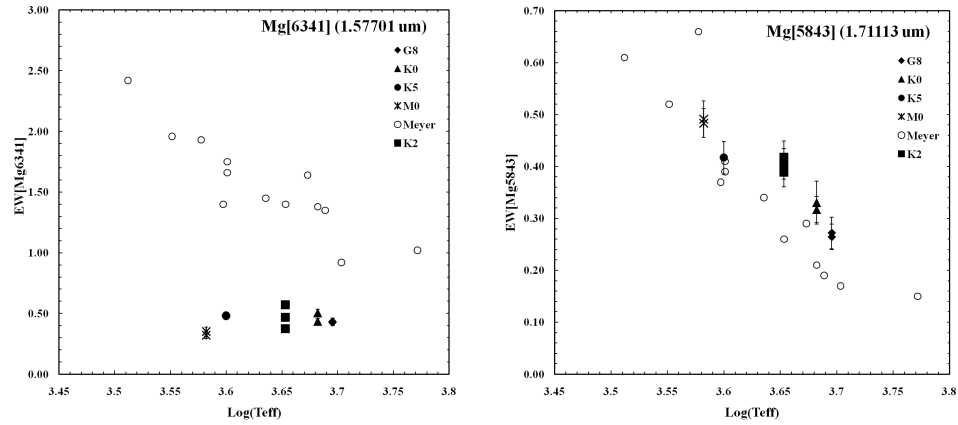}
\caption{EWs of Mg (1.57701 $\mu$m) and Mg (1.71113 $\mu$m). Mg (1.57701 $\mu$m) is contaminated with other lines, but Mg (1.71113 $\mu$m) is a single line. Our results are smaller than M98 results because of the different spectral resolution. The width ranges used for calculated EWs in M98 may be wider than ours, and made the results different. This we can be seen in comparison of EWs of two Mg lines of Arcturus spectrum with different resolution. The single Mg (1.71113 $\mu$m) line is very sensitive to the effective temperature, and we use this line to estimate effective temperature of late-type stars.}
\label{Mg2lines}
\end{figure*}

\begin{figure}[h!]
\centering
\includegraphics[width=45ex]{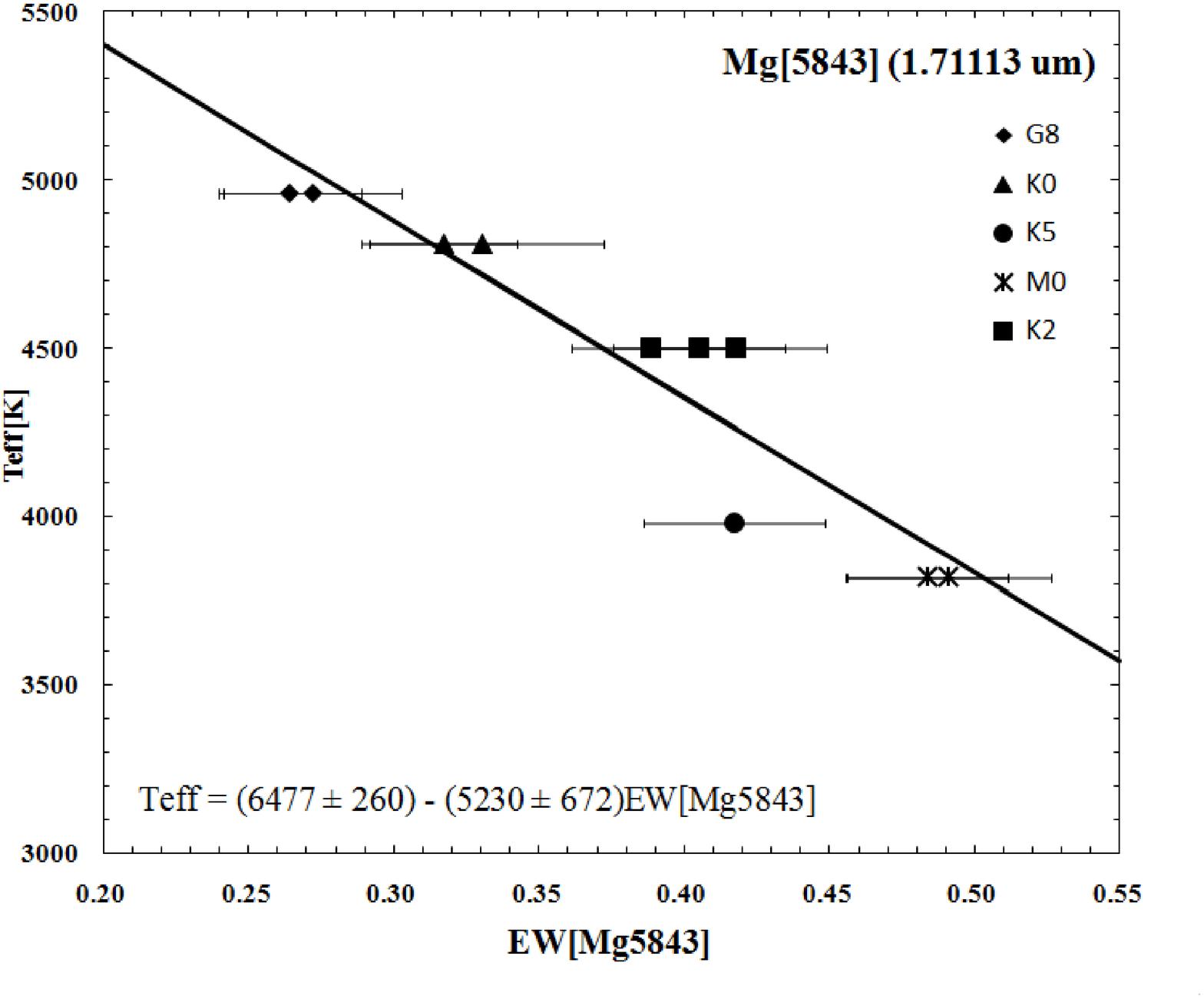}
\caption{Linear fitting of EWs of Mg(1.71113 $\mu$m) vs. effective temperature. The error bars indicate the errors of EW for each star that have been calculated from formula (2).}
\label{linear}
\end{figure}

We present 4 specific lines which show the different relation between effective temperature and EWs. Fig. \ref{allline}
shows the lines of Si (1.59644 $\mu$m), Si (1.66853 $\mu$m), Mg (1.57701 $\mu$m) and Mg (1.71113 $\mu$m) in our spectra of G8-M0 stars, including Arcturus (K2 III). The results of EWs and effective temperatures are presented in Fig. \ref{Si2lines} and \ref{Mg2lines}. From these figures, we can see that our EW results are smaller than those of M98. This discrepancy of EWs might come from different spectral resolution between our spectra and that of M98.

In M98, the authors might have calculated the EWs with wider range than ours. In this case, the EW results might be contaminated with nearby lines. As shown in Fig. \ref{Si2lines}, the EW trend of Si (1.59644 $\mu$m) line in our result is different from that in M98. It is likely that the EW of Si (1.59644 $\mu$m) line in M98 includes the strong line of Fe (1.59851 $\mu$m), which is increasing with decreasing effective temperature, as seen in Fig.
\ref{allline}.

In the case of Si (1.66853 $\mu$m), we can see that it has also many nearby lines which show increasing trend of EW with decreasing effective temperature, as shown in Fig.
\ref{allline}.
Therefore, it is obvious that the EW trends of Si lines in M98 are affected by nearby lines which are increasing with decreasing effective temperature.

Fig. \ref{Mg2lines} shows the different EW trends of Mg (1.57701 $\mu$m) between our result and M98. It is similar to the case of the EW of Si (1.59644 $\mu$m) line in M98. In case of Mg (1.57701 $\mu$m), M98 calculation might include the strong nearby line of 1.57800 $\mu$m which is increasing with decreasing effective temperature.

In Fig. \ref{Mg2lines}, we also find the fact that the Mg (1.71113 $\mu$m) line is very sensitive to the temperature of giant stars in G, K, M type, rather than Mg (1.57701 $\mu$m) which has been mentioned in M98. The linear fitting of the EWs of Mg line is shown in Fig. \ref{linear}. Using the EW trend of Mg (1.71113 $\mu$m) line for effective temperature, therefore, the relation between effective temperature and EW of Mg line could be obtained by linear fitting. We find the equation to estimate effective temperature from the EW of Mg (1.71113 $\mu$m) line as
\begin{equation}
T_{eff} = 6447 \pm 260 - (5230 \pm 672) EW[Mg(5843)]
\end{equation}
The standard deviation error of the slope in the linear fitting is large because of the low number of stars. This formula can be used to estimate the effective temperature of late-type giant stars (G8 III-M0 III) from EW measurement.

\section{SUMMARY AND FUTURE WORKS}

We present the method for data analysis of the near-IR medium resolution echelle spectrometer, IRCS, at the SUBARU 8.2m telescope. From the method, we have shown the medium resolution template star spectra of HD64938 and HD148287 (G8 III), HD55184 and HD155500 (K0 III), HD52071, HD146084, and HD122675 (K2 III), HD154610 (K5 III), and HD76010 and NSV3729 (M0 III) in H-band.

We found many prominent molecular and neutral metal features in the template star spectra. We measured the EWs of the features and compare our results to M98 results. Our results with spectral resolution (R $=$ $6000$) are more accurate than M98 with spectral resolution (R $=$ $3000$). We use the Mg(1.71113 $\mu$m) line to estimate approximate temperature for late-type stars.

In this work, the obtained H-band spectra of late-type stellar templates would play important roles for studying relation between active galactic nuclei (AGNs) and their host galaxy properties. The broad absorption $\nu$ $=$ $3-6$ $^{12}$CO band head at 1.62073 $\mu$m obtained from the spectra is potentially useful for measuring the velocity dispersion in the central region of the QSO host galaxies. The library of the template stellar spectra in ASCII format can be downloaded from the website http:$\slash$$\slash$irlab.khu.ac.kr$\slash$$\sim$anh$\slash$file$\slash$StellarData.zip

\acknowledgments
{This work was supported by the National Research Foundation of Korea (NRF) grant funded by the Korean government (MEST), No. 2009-0063616.} \\


\end{document}